\documentclass[aps,prb,twocolumn,showpacs,superscriptaddress,floatfix]{revtex4}
\usepackage{amsfonts}
\usepackage{amsmath}
\usepackage{amssymb}
\usepackage{graphicx}
\begin{document}
\title{Scaling of the anomalous Hall effect in SrRuO$_{3}$}
\date{\today }

\author{Noam Haham}
\author{Yishai Shperber}
\author{Moty Schultz}
\author{Netanel Naftalis}
\author{Efrat Shimshoni}
\affiliation{Department of Physics, Nano-magnetism Research Center, Institute of Nanotechnology and Advanced Materials, Bar-Ilan University,
Ramat-Gan 52900, Israel}
\author{James W. Reiner}
\affiliation{Hitachi Global Storage Technologies, 3403 Yerba Buena Rd, San Jose, CA 95315}
\author{Lior Klein}
\affiliation{Department of Physics, Nano-magnetism Research Center, Institute of Nanotechnology and Advanced Materials, Bar-Ilan University,
Ramat-Gan 52900, Israel}

\begin{abstract}

We measure the anomalous Hall effect (AHE) resistivity $\rho_{xy}$ in thin films of the itinerant ferromagnet SrRuO$_{3}$.
At low temperatures, the AHE coefficient $R_{s}$ varies with $\rho_{xx}^2$ ,and at higher temperatures, $R_{s}$ reaches a peak and then changes sign just below $T_{c}$. We find that for all films studied $R_{s}$ scales with resistivity in the entire ferromagnetic phase. We attribute the observed behavior to the contribution of the extrinsic side jumps mechanism and the intrinsic Karplus-Luttinger (Berry phase) mechanism including the effect of finite scattering rates.

\end{abstract}

\maketitle


\section{introduction}
Being one of the most intriguing manifestations of a transport phenomenon that is sensitive to spin and topology, the anomalous Hall effect (AHE) \cite{AHE} is at the focus of considerable theoretical and experimental efforts.
The interest in spin sensitive phenomena is linked to the emerging field of spintronics \cite{spintronics}, which offers an alternative
to conventional charge-based electronics. The interest in the effects of topological features of bands on transport properties is linked to the role that these effects play in systems such as topological insulators and quantum Hall systems \cite{topological insulators}.

The AHE is described phenomenologically as transverse resistivity $\rho^{AHE}_{xy}$ or transverse conductivity
$\sigma^{AHE}_{xy}$ linked to the intrinsic magnetization $\vec{M}$ of
a conductor. Various models have been proposed: (a)
The extrinsic model  relates the AHE to antisymmetric scattering
processes and it provides that
 \begin{equation}
 \rho_{xy}^{AHE}=R_s\mu_0M_\bot
 \label{EqRs}
  \end{equation}
  where $R_{s}=a\rho_{xx}+b\rho_{xx}^{2}$ and $M_\bot$ is the component of magnetization perpendicular to the film. The linear term in resistivity of $R_{s}$ is attributed to
 skew scattering \cite{EHE Smit} and it is expected to dominate in high conductivity regime ($\sigma_{xx}>10^{6} \ \Omega^{-1} \ {\rm cm}^{-1}$). The quadratic term is attributed to side jumps \cite{EHE Berger} and it is expected to dominate in the good conductivity regime ($\sigma_{xx}\sim10^{4}- 10^{6} \ \Omega^{-1} \ {\rm cm}^{-1}$).
(b) The intrinsic model known also as the Karplus-Luttinger model (K-L) \cite{Luttinger} or Berry phase model attributes the AHE to intrinsic topological properties
of the band \cite{BerryPhase1,BerryPhase2}.
According to this model
 $\rho_{xy}^{AHE}=\rho_{xx}^2 \sigma_{xy}(\vec{M})$ and it is expected to dominate in the same regime as the side jump mechanism.
 In the poor conductivity regime ($\sigma_{xx} < 10^4 \ \Omega^{-1} \ {\rm cm}^{-1}$) a universal behavior  $\sigma_{xy} \sim \sigma_{xx}^{1.6-1.8}$ has been observed experimentally \cite{bad metal scaling experiment}; however, a theoretical understanding is still lacking. Interestingly, a similar scaling is predicted for metals in the limit of strong scattering due to finite-lifetime disorder broadening \cite{bad metal scaling theory}, and within a microscopic model accounting for fluctuations of local orbital energies \cite{streda}.

SrRuO$_{3}$ has played a pivotal role in the study of the AHE and numerous attempts have been made to elucidate its complicated behavior. Berry phase calculations which assume a temperature-dependent exchange gap that closes at $T_{c}$ seemed to describe the data reasonably \cite{BerryPhase2}. However, a test of this scenario that focused on the vanishing point of the AHE found that it vanishes for a given film (whose resistivity and magnetization were varied by field) at a specific resistivity, and not at a specific magnetization as one may expect from a scenario which attributes the vanishing signal of the AHE to the Berry phase contribution at a particular exchange splitting \cite{berryphase}. Mid infrared measurements suggest the applicability of the Berry phase scenario at energies above 200 meV while the $dc$ limit is dominated by extrinsic scattering mechanisms \cite{infrared}.

By using SrRuO$_{3}$ films with a wide range of thicknesses that vary considerably in the temperature-dependence of their resistivity, we provide a compelling piece of evidence that resistivity, \emph{irrespective of its sources or nature (elastic or inelastic)}, determines the AHE of SrRuO$_{3}$ in the entire ferromagnetic phase.  This observation strongly suggests that changes in Berry phase due to assumed temperature-dependent exchange splitting cannot explain the complicated temperature dependence of the AHE. We show that the side jumps mechanism combined with the Karplus-Luttinger (Berry phase) mechanism that takes into account the scattering time may explain the observed behavior.
\begin{figure}[ptb]
\includegraphics[scale=0.6, trim=100 200 0 100] {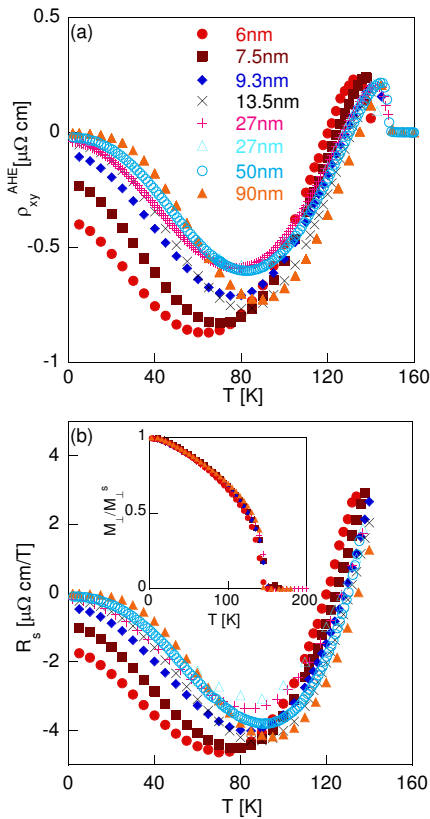}
\caption{(a) Remanent AHE resistivity ($\rho_{xy}^{AHE}$) of 8 films vs temperature. (b)  The AHE coefficient ($R_{s}$) vs temperature derived from $\rho_{xy}^{AHE}$ and $M_\bot$ using Eq. 1 . Inset: scaling of the perpendicular magnetization ($M_{\bot}$) normalized by its low temperature value as a function of temperature for 5 films with thickness between 6 to 90 nm.}
 \label{Layout1}
\end{figure}

\section{samples and experiment}

Our samples are epitaxial thin films of SrRuO$_{3}$
grown on slightly miscut ($\sim{2}^{\circ}$) substrates of SrTiO${_3}$
by reactive electron beam evaporation. The films are untwinned orthorhombic single-crystals, with lattice parameters of $a\cong5.53${\AA}, $b\cong5.57${\AA}, and $c\cong7.85${\AA}.  The films were patterned to allow transverse and longitudinal resistivity measurements, which were performed with a Quantum Design PPMS-9. The films exhibit exceptionally high resistivity ratio (up to 90) indicative of their high quality. The thinnest films ($\leq 10\ \rm{nm}$) exhibit lower resistivity ratio ($\ge 5$) which is still very high considering the enhanced surface scattering. Magnetic characterization of the films was performed using a Quantum Design SQUID magnetometer (MPMS).

Magnetic films may exhibit AHE if their magnetization has a component perpendicular to the film plane. As shape anisotropy favors in-plane magnetization, in many cases a perpendicular field should be applied in order to tilt the magnetization out of the plane. This may complicate the analysis since the applied field also induces ordinary Hall effect (OHE).

SrRuO$_{3}$ films exhibit intrinsic uniaxial magnetocrystalline anisotropy with an easy axis which varies with temperature between ~45 degrees to the normal at $T_{c}$ to ~30 degrees at 2 K \cite{reorientation}. Moreover, the remanent magnetization is stable and spontaneous breakdown into magnetic domains occurs only a few degrees below $T_{c}$ \cite{stable magnetization}. These features enable direct measurement of zero field (remanent) antisymmetric transverse resistivity which can be fully attributed to AHE, $\rho_{xy}^{AHE}$.

\begin{figure}[ptb]
\includegraphics[scale=0.6, trim=100 200 0 100] {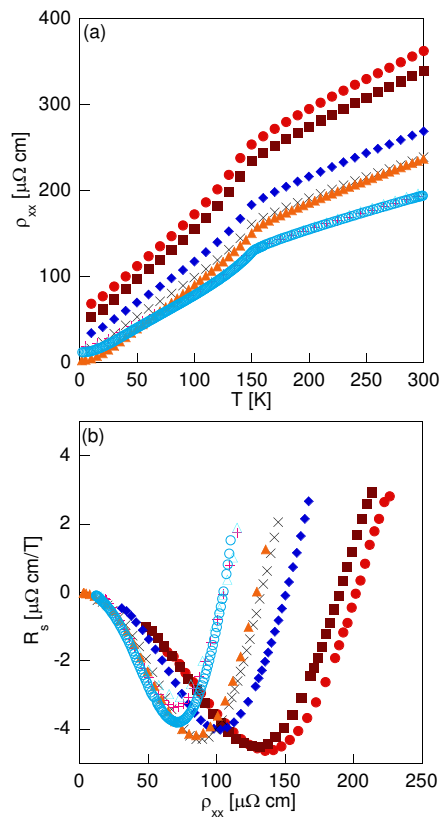}
\caption{(a) Longitudinal resistivity ($\rho_{xx}$) of the 8 films presented in Fig.$~\ref{Layout1}$, vs temperature. (b) $R_{s}$ from Fig.$~\ref{Layout1}$(b) vs resistivity ($\rho_{xx}$).}
 \label{Layout2}
\end{figure}

\section{experimental results}

Fig.~$\ref{Layout1}$(a) shows temperature dependence of $\rho_{xy}^{AHE}$ of 8 different samples. To extract $R_{s}$ based on Eq. 1 (see Fig.~\ref{Layout1}(b)), we divide $\rho_{xy}^{AHE}$ by $M_{\bot}$ (shown in the inset). We note that for the thickness range of our samples, $M_{\bot}$ is practically identical except for small deviations related to thickness-dependent $T_{c}$. This is expected as thickness-induced changes in magnetic properties were reported for films with thickness lower than 6 nm \cite{moty2}. We note that while $R_{s}$ of the various samples has general common features, the variations are considerable. In particular we note differences in the values of $R_{s}$ at 2 K, in the location of the negative peak and in the temperature at which $R_{s}$ changes its sign. The large spread in $R_{s}$ seems to correlate with changes in the resistivity of the films, strongly affected by film thickness (see Fig.~$\ref{Layout2}$(a)). However, as seen in Fig.~$\ref{Layout2}$(b), the extracted
$R_s$ does not scale with $\rho_{xx}$. In particular, we note that the resistivity at which $R_s$ changes its sign ($\rho_{0}$) varies between 105 $\mu\Omega \ {\rm cm} $ for a 50 nm thick sample to 202 $\mu\Omega \ {\rm cm} $ for a 6 nm thick sample. Does this observation exclude the scenario that $R_s$ is determined by $\rho_{xx}$ in the entire ferromagnetic phase? - not necessarily.

\begin{figure}[ptb]
\includegraphics[scale=0.5, trim=150 0 120 0]{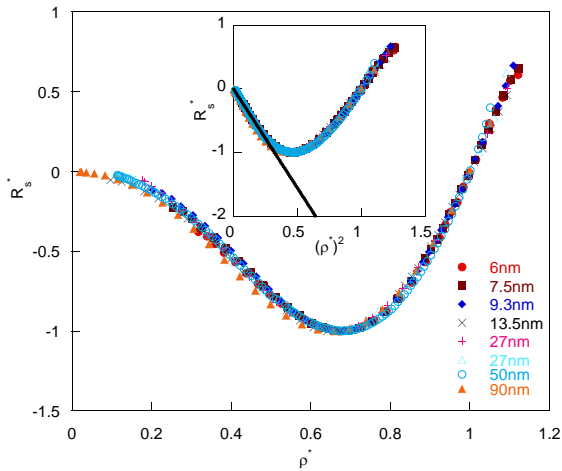}
\caption{$R_{s}$ normalized by its absolute maximum value (${R_{s}}^{*}$) vs $\rho_{xx}$ normalized by its value when $R_{s}$ changes its sign ($\rho^{*}$). Inset: ${R_{s}}^{*}$ vs $({{\rho}^{*}})^{2}$.}
 \label{Layout3}
\end{figure}

Fig.~$\ref{Layout3}$ shows that $R_{s}^{*}$, defined as $R_{s}$ normalized by its maximum absolute value, does scale with $\rho^{*}$, defined as $\rho_{xx}$ normalized by $\rho_{0}$.
The scaling function has a quadratic dependence on ${\rho}^{*}$ in the low resistivity regime (see inset) and it reaches its negative peak for all samples at ${\rho}^{*}\cong0.7$.

\begin{figure}
\includegraphics[scale=0.4, trim=100 0 100 0]{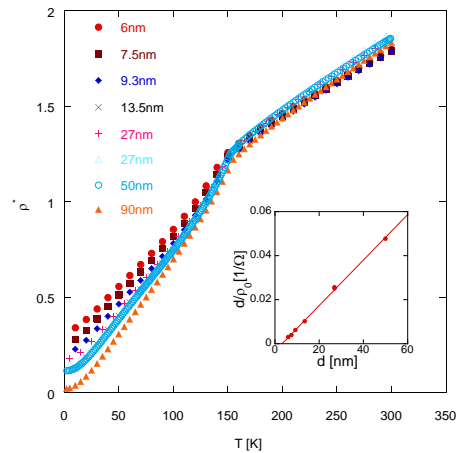}
\caption{Normalized resistivity (${\rho}^{*}$) vs temperature. Inset: film thickness (d) divided by its resistivity when $R_{s}$ changes its sign ($\rho_{0}$) vs film thickness (d).}
 \label{Layout4}
\end{figure}

A possible explanation for the striking scaling is that $R_s$ is determined by $\rho_{xx}$ and that it \emph{does} vanish at the same intrinsic resistivity $\rho_{0}^{int}$ for all samples, consistent with a previous report \cite{berryphase}; however, there is a multiplicative factor $\gamma$ between the nominal resistivity and the intrinsic resistivity, $\rho_{xx}=\gamma\rho_{xx}^{int}$. A trivial source for $\gamma$ is uncertainty in film thickness and in geometrical factors of the pattern. However, these sources alone cannot account for the observed variations of order 2. Another potential source is dead layers \cite{moty2, deadlayers} whose existence may affect considerably the calculated resistivity of ultrathin films. Assuming a dead layer of thickness $\delta$, we would expect $\gamma=d/(d-\delta)$ and a linear dependence between $d/\rho_{0}$ and $d$, as observed in the inset of Fig.~$\ref{Layout4}$. The linear fit is consistent with a dead layer scenario with $\delta\sim 3 \ {\rm nm}$  and $\rho_{0}^{int}\sim 100 \ \mu\Omega \ {\rm cm}$. The dead layer scenario also implies that the resistivity of the various samples at high temperatures is not different (as suggested by Fig.~$\ref{Layout2}$(a)) but quite similar (Fig.~$\ref{Layout4}$). As the main difference between the films is in their thickness, the result supports the dead layer scenario as it is expected that at high temperatures, where the mean free path is small and bulk scattering is dominant, the resistivity of our films would be similar.

The dead layer scenario implies the need to normalize $R_{s}$; However, its division by $\gamma$ does not scale the data along the \emph{y} axis. Therefore, the normalization of $R_{s}$ with its maximum absolute value merely indicates that for all films there is a single $R_{s}(\rho_{xx})$ function up to a multiplicative factor.

We note that the scaling is obtained for films that vary considerably in their thickness and residual resistivity; namely the same value of $R_s$ is obtained for very different values of $M_\bot$ and for very different contributions to $\rho_{xx}$.
Thus for instance, $R_s$ attains its maximum value at $T/T_{c}$=0.47 for the 6 nm thick film and at $T/T_{c}$=0.63 for the 90 nm thick film. At this temperature the magnetization is 84 percent (77 percent) of its low temperature value for the thin thick) film  and the
resistivity is 2 times (32 times) larger than its low temperature value. Therefore, point defects, surface scattering, magnons and phonons have very different weights in the two cases.

\section{theoretical model}

 The low temperature dependence of R$_s$ on $\rho_{xx}^2$ is consistent with side jumps mechanism \cite{EHE Berger} and with the Karplus-Luttinger (K-L) or Berry phase mechanism \cite{Luttinger}. However, whereas side jumps can explain the scaling with $\rho_{xx}$ due to its insensitivity to the scattering potential, it cannot explain the non monotonic temperature dependence which includes a sign change at higher temperatures. On the other hand, attributing the non monotonic temperature dependence to K-L mechanism with temperature dependent exchange gap as suggested previously \cite{BerryPhase2}, yields $\rho_{xy}^{AHE}=\rho_{xx}^2\sigma_{xy}(\textbf{M})$ with a complicated dependence of $\sigma_{xy}$ on $\textbf{M}$ which is inconsistent with the scaling which assumes linear dependence on $M_\bot$.

We now show that a combination of the side jumps mechanism and the K-L mechanism which considers the effect of scattering rate and its temperature dependence (without assuming any change in the band structure) is a possible scenario. A consideration of the scattering rate ($1/\tau$) effect on the transverse conductivity in the K-L mechanism is required in moderately good conductors, where $\hbar/\tau$ is not negligible compared to the inter-band gap. The leading correction
yields a decrease of $\sigma_{xy}$ as the resistivity increases, and thus a possible non-monotonic behavior of $\rho_{xy}$.

Accounting for a finite $\tau$, the K-L contribution to the AHE resistivity from Kubo's formula \cite{Kubo} becomes
 \begin{equation}
 \begin{array}{ll}
 \rho^{K-L}_{xy}=\rho^2_{xx}e^{2}\hbar/\Omega \\ \sum_{n\neq m,k}\frac{<n k|v_{y}|m k><m k|v_{x}|n k>(f(\varepsilon_{n,k})-f(\varepsilon_{m,k}))}{\{i(\varepsilon_{m,k}-\varepsilon_{n,k})+\hbar/\tau\}\{\varepsilon_{n,k}-\varepsilon_{m,k}\}}
 \label{Eqrhoxy}
 \end{array}
  \end{equation}
where $\Omega$ is the crystal volume, $k$ is the quasi-momentum $n,m$ are band indices associated with the eigenvalues of the perfect crystal Hamiltonian, $v_x$, $v_y$ are the velocity operators, and $f(\varepsilon)$ is the Fermi-Dirac distribution. This contribution accounts only for the intrinsic part of the AHE, i.e. ignores the corrections to the scattering processes due to spin-orbit interaction.
 \begin{figure}
\includegraphics[scale=0.4, trim=100 0 100 0]{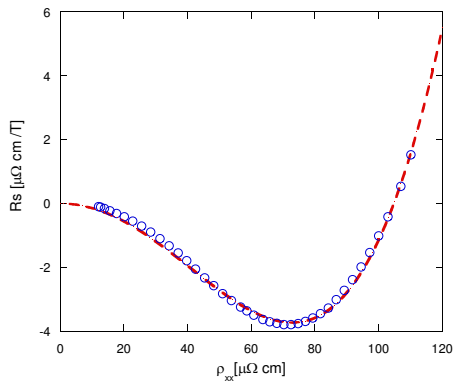}
\caption{AHE coefficient R$_s$ as a function of resistivity $\rho_{xx}$ for a thick film (500 \AA). The dashed line is a fit to Eq. 8.}
 \label{Layout5}
\end{figure}
We consider a model in which the main contribution to the sum in Eq. \ref{Eqrhoxy} is due to two bands denoted as 1,2, where the Fermi level crosses the upper band while the lower band is fully occupied. We further assume that the dominant contribution arises from states with quasi-momentum in a set denoted as $\mathbb{K}$, and that the energy gap for $k\in\mathbb{K}$ between non-occupied states in the upper level and occupied states in the lower level is approximately independent of quasi-momentum and takes the characteristic value of $\Delta$. Under these assumptions, we obtain
 \begin{equation}
 \rho^{K-L}_{xy}=-\rho^2_{xx}e^{2}\hbar(\frac{A}{\{i\Delta+\hbar/\tau\}\Delta}+\frac{A^*}{\{-i\Delta+\hbar/\tau\}\Delta})
 \label{4}
 \end{equation}
 where $A$ is defined as
 \begin{equation}
 A\equiv\int_{\mathbb{K}}\frac{d^3k}{{2\pi}^3}<1 k|v_{y}|2 k><2 k|v_{x}|1 k>
 \end{equation}
 which can be associated with a Berry's phase \cite{AHE}. Since
 $A$ is odd under time-reversal and hence purely imaginary, we get
  \begin{equation}
 \rho^{K-L}_{xy}=-\rho^2_{xx}e^{2}\hbar\frac{2Im(A)}{\Delta^2+(\hbar/\tau)^2}
 \label{4}\; .
 \end{equation}
Considering spin orbit interaction (SOI), A is expected to be proportional to $M_\bot$ \cite{Luttinger}; namely,
  \begin{equation}
 Im(A)=aM_\bot
 \end{equation}
 where $a$ is a constant.
 Previous reports indicate that the band structure in SrRuO$_3$ is temperature independent \cite{splitting}, thus, ferromagnetism in SrRuO$_3$ should be described in the local band model \cite{local band}. Therefore considering Eqs. 5 and 6 and preforming averaging yield:
 \begin{equation}
 \rho^{K-L}_{xy}=-\rho^2_{xx}e^{2}\hbar\frac{2a}{\Delta^2+(\hbar/\tau)^2}M_\bot
 \label{5}\;
 \end{equation}
 where $M_\bot$ is the averaged magnetization in the sample.
 Finally, we note that within the same level of approximation (i.e., leading order in the scattering potential), the side jumps contribution is additive to the K-L term \cite{diagrams}. Thus, the AHE coefficient (R$_s$) is given by a sum of the two contributions:
   \begin{equation}
 R_s=\rho^2_{xx}\frac{B}{\Delta^2+(\hbar/\tau)^2}+C\rho^2_{xx}
 \label{5}\; .
 \end{equation}
The first term is the K-L term (with all the constants and the minus sign included in $B$), and the second term is the side jumps contribution. As $B$, $\Delta$ and $C$ are merely associated with the band structure they are assumed to be constants. $1/\tau$ is assumed to be, as usual, proportional to $\rho_{xx}$ and the proportionality factor is estimated based on band calculations \cite{band}. Thus, the right hand side in Eq. 8 is a function of $\rho_{xx}$ alone.

Fig.~$\ref{Layout5}$ shows a fit of our data using Eq. 8 where the parameter C is limited to an interval which corresponds to a reasonable range of side jumps (0.1-10 \AA ~\cite{EHE Berger}). We obtain a good fit for side jumps in the range of 1-10 \AA, and $\Delta$ in the range of 0.07-0.2 eV. The value of $\Delta$ is in good agreement with the characteristic energy at which Im($\sigma_{xy}$) has a peak, measured in the infrared regime for the low temperature limit \cite{infrared}.
The fit presented in Fig.~$\ref{Layout5}$ is for a side jump $\sim$ 4 \AA ~ and $\Delta$ $\sim$ 0.13 eV. The fact that a similar temperature dependence of R$_s$ is observed for other systems \cite{other materials} suggests that this scenario is relevant to other materials as well.

\section{conclusions}
The scaling of  $\rho_{xy}^{AHE}$ data with $\rho_{xx}$ in SRO films implies that the AHE coefficient is determined by the total resistivity irrespective of the relative contributions of different scattering processes. To explain the scaling and the non monotonic behavior of the scaling function we present a scenario that attributes the observed behavior to two contributions: (a) side jumps mechanism and (b) K-L (Berry phase) mechanism including the effect of finite scattering rates. In the limit of low resistivity, the two contributions have quadratic dependence on resistivity with coefficients of opposite signs where that of K-L term is larger. As resistivity increases, the K-L term decays due to the effect of finite scattering rates which yields a sign change of  $\rho_{xy}^{AHE}$.

\section{acknowledgments}
We acknowledge useful discussions with J. S. Dodge, Y. Kats and S. Simon.
L.K. acknowledges support by the Israel Science Foundation founded by the Israel Academy of Sciences and Humanities
(Grant 577/07). E. S. acknowledges support by the Israel Science Foundation (Grant 599/10), the US--Israel Binational Science Foundation (Grant No. 2008256) and the Aspen Center for Physics.
J.W.R. grew the samples at Stanford University in the laboratory of M.R. Beasley.


\begin{thebibliography}{2}

\bibitem {AHE} N. Nagaosa, J. Sinova, S. Onoda, A. H. MacDonald, and N. P. Ong, Rev. Mod. Phys. \textbf{82}, 1539 (2010).

\bibitem {spintronics} S. D. Bader and S. S. P. Parkin,
 Annu. Rev. Condens. Matter Phys. \textbf{1}, 71 (2010).

\bibitem {topological insulators} X. Qi and S. Zhang, Phys. Today \textbf{63 (1)}, 33 (2010).

\bibitem {EHE Smit} J. Smit, Physica \textbf{21}, 877 (1955);
J. Smit, Physica \textbf{24}, 39 (1958).

\bibitem {EHE Berger} L. Berger, Phys. Rev. B
\textbf{2}, 4559 (1970);
L. Berger, Phys. Rev. B
\textbf{5}, 1862 (1972).

 \bibitem{Luttinger} R. Karplus and J. Luttinger, Phys. Rev. \textbf{95}, 1154 (1954).
\bibitem {BerryPhase1} T. Jungwirth, Q. Niu, and A. H. MacDonald, Phys. Rev. Lett. \textbf{88}, 207208 (2002); Y. Yao, L. Kleinman, A. H. MacDonald, J. Sinova, T. Jungwirth, D.-S. Wang, E. Wang, and Q. Niu, Phys. Rev. Lett. \textbf{92}, 037204 (2004).

\bibitem {BerryPhase2}  Z. Fang, N. Nagaosa, K. S. Takahashi, A. Asamitsu, R. Mathieu, T. Ogasawara, H. Yamada, M. Kawasaki, Y. Tokura, and K. Terakura, Science \textbf{302}, 92 (2003); R. Mathieu, C. U. Jung, H. Yamada, A. Asamitsu, M. Kawasaki, and Y. Tokura, Phys. Rev. B \textbf{72}, 064436 (2005).

\bibitem {bad metal scaling experiment} T. Fukumura, H. Toyosaki, K. Ueno, M. Nakano, T. Yamasaki, and M. Kawasaki, Jpn. J. Appl. Phys. \textbf{46}, L642 (2007); D. Venkateshvaran, W. Kaiser, A. Boger, M. Althammer, M. S. Ramachandra Rao, S. T. B. Goennenwein, M. Opel, and R. Gross, Phys. Rev. B \textbf{78}, 092405 (2008); W. R. Branford, K. A. Yates, E. Barkhoudarov, J. D. Moore, K. Morrison, F. Magnus, Y. Miyoshi, P. M. Sousa, O. Conde, A. J. Silvestre, and L. F. Cohen, Phys. Rev. Lett. \textbf{102}, 227201 (2009); T. Miyasato, N. Abe, T. Fujii, A. Asamitsu, S. Onoda, Y. Onose, N. Nagaosa, and Y. Tokura, Phys. Rev. Lett. \textbf{99}, 086602 (2007).

\bibitem {bad metal scaling theory} S. Onoda, N. Sugimoto, and N. Nagaosa, Phys. Rev. B \textbf{77}, 165103 (2008); S. Onoda, N. Sugimoto, and N. Nagaosa, Phys. Rev. Lett. \textbf{97}, 126602 (2006).

\bibitem {streda}
P. Streda, Phys. Rev. B \textbf{82}, 045115 (2010).

\bibitem {berryphase} Y. Kats, I. Genish, L. Klein, J. W. Reiner, and M. R. Beasley, Phys. Rev. B \textbf{70}, 180407 (2004).

\bibitem {infrared} M.-H. Kim, G. Acbas, M.-H. Yang, M. Eginligil, P. Khalifah, I. Ohkubo, H. Christen, D. Mandrus, Z. Fang, and J. Cerne, Phys. Rev. B \textbf{81}, 235218 (2010).

\bibitem {reorientation} L. Klein, J. S. Dodge, C. H. Ahn, J. W. Reiner, L. Mieville, T. H. Geballe, M. R. Beasley, and A. Kapitulnik, J. Phys. Condens. Matter \textbf{8}, 10111 (1996).

\bibitem {stable magnetization} A. F. Marshall, L. Klein, J. S. Dodge, C. H. Ahn, J. W. Reiner, L. Mieville, L. Antagonazza, A. Kapitulnik, T. H. Geballe, and M. R. Beasley, J. Appl. Phys. \textbf{85}, 4131 (1999).


\bibitem {moty2}  M. Schultz, S. Levy, J. W. Reiner, and L. Klein,
Phys. Rev. B \textbf{79}, 125444 (2009).
\bibitem {deadlayers} R. P. Borges, W. Guichard, J. G. Lunney, J. M. D. Coey, and F. Ott, J. Appl. Phys. \textbf{89}, 3868 (2001); J. Z. Sun, D. W. Abraham, R. A. Rao, and C. B. Eom, Appl. Phys. Lett. \textbf{74}, 3017 (1999).

 \bibitem{Kubo} Y. Murayama, \emph{Mesoscopic Systems: Fundamentals  and Applications} pp.213 (App. G).
 \bibitem {splitting} J. S. Dodge, E. Kulatov, L. Klein, C. H. Ahn, J. W. Reiner, T. H. Geballe, M. R. Beasley, A. Kapitulnik, H. Ohta, Yu Uspenskii, and S. Halilov, Phys. Rev. B \textbf{60}, R6987 (1999).
 \bibitem{local band} V. Korenman, J. L. Murray, and R. E. Prange, Phys. Rev. B \textbf{16}, 4032 (1977);
V. Korenmanand and R. E. Prange, Phys. Rev. Lett. \textbf{53}, 186 (1984).
 \bibitem{diagrams}
 H. Kontani, T. Tanaka and K. Yamada, Phys. Rev. B \textbf{75}, 184416 (2007); A. Crepieux and P. Bruno, Phys. Rev. B \textbf{64}, 014416 (2001).
\bibitem{band} G. Santi and T. Jarlborg, J. Phys.: Condens. Matter \textbf{9}, 9563 (1997).
\bibitem{other materials} J. G. Checkelsky, M. Lee, E. Morosan, R. J. Cava, and N. P. Ong, Phys. Rev. B \textbf{77}, 014433 (2008).


\end{thebibliography}
\end{document}